\documentstyle[prd,epsf,aps]{revtex}

\def\bra#1{\mathopen{\langle#1\,|}}
\def\ket#1{\mathclose{|\,#1\rangle}}
\def\lrarr#1{\vbox{\ialign{##\crcr
    $\leftrightarrow$\crcr
    \noalign{\kern 1pt\nointerlineskip}
    $\hfil\displaystyle{#1}\hfil$\crcr}}}
\def\larr#1{\vbox{\ialign{##\crcr
    $\leftarrow$\crcr
    \noalign{\kern 1pt\nointerlineskip}
    $\hfil\displaystyle{#1}\hfil$\crcr}}}
\def\rarr#1{\vbox{\ialign{##\crcr
    $\rightarrow$\crcr
    \noalign{\kern 1pt\nointerlineskip}
    $\hfil\displaystyle{#1}\hfil$\crcr}}}

\title{Gerasimov--Drell--Hearn sum rule and nucleon structure}
\author{S.Ying}
\address{
Physics Department, Fudan University\\ Shanghai 200433,
China}
\date{\today}

\tighten
\draft

\begin{document}

\maketitle

\begin{abstract}
{ A modification of the Gerasimov--Drell--Hearn sum rule suggested by
   the current experimental data is presented. Within the conventional
   theoretical framework, we find it necessary to consider the
   possibility of the presence of a localized region inside a nucleon,
   in which the electromagnetic (EM) gauge symmetry is spontaneous
   broken down, if the constraints of the gauge invariance, Lorentz
   invariance and the assumption of the commutativity of the EM charge
   density operator at equal-time are considered. We also discuss the
   propagation of a virtual photon inside a nucleon under such a
   scenario.  A comment on some of the recent model independent works
   on the same subject is provided.}
\end{abstract}


\section{Introduction}

   There are a few sum rules from current algebra that can be
subjected to experimental tests. The Gerasimov--Drell--Hearn (GDH) sum
rule \cite{GMV,DH}, which can be viewed as a superconvergence
condition that depends only on the large energy behavior of the
scattering amplitude \cite{AdlerBook}, is one of the well founded
ones. It is related to the spin structure of a nucleon. Recent
analysis \cite{WA,SWK} of the currently available experimental data
for the photo-production of pion on a nucleon below the photon energy
of 1.7 GeV suggests, however, a violation of the sum rule.  If this
violation is proven genuine in the future, it would imply that our
understanding of the structure of a nucleon is probably incomplete.
At the present, it is however still unclear whether the violation is
due to the inaccuracy of experimental information or is genuine in
nature. This is because the result obtained in Refs. \cite{WA,SWK} are
constructed from single pion photoproduction data (together with an
estimation of two pion photoproduction contributions) from different
charge channels with an upper energy cutoff at 1.7 GeV. Albeit the
results seem to be quite stable, they still need to be confirmed by
direct measurements of total photon--nucleon cross section
$\sigma_{3/2}$ and $\sigma_{1/2}$, which correspond to the ones with
photon helicity parallel and antiparallel to the nucleon spin
polarization respectively.  In the similar energy region, experiments
in that direction are planed at the BNL Laser Electron Gamma Source
(LEGS), Mainz (MAMI) accelerator facilities. In higher energy regions,
they are planed at the Continuous Electron Beam Accelerator Facility
(CEBAF), the Bonn Electron Stretcher Accelerator (ELSA) and the
Grenoble Anneau Accelerator Laser (GRAAL) facility.

  By supposing that the above mentioned violation is a possibility,
theoretical investigations on how such a violation be explained by an
extension of the GDH sum rule can be carried out. The extent to which
the GDH sum rule should be extended and what such an extension implies
should ultimately be determined by observations. Nevertheless, several
schemes were already provided based on 1) the assumption of the
existence of a fixed J=1 pole in the helicity amplitude of the
photon--nucleon Compton scattering \cite{AG,FGFD} and 2) the
modification of the commutator between electromagnetic (EM) charge
density operators at equal-time \cite{KS,CLW}. Since in these schemes,
the requirements of Lorentz and local EM gauge invariance are not
considered explicitly, a third one is provided in this paper by
imposing the constraints of these requirements.  Through this study,
an extension of the GDH sum rule based on the possibility of a
(localized) spontaneous breaking down of the $U(1)$ gauge symmetry
corresponding to the electromagnetism is proposed.

   In section \ref{sec:G-inv}, the condition of gauge invariance on
the photon--nucleon forward Compton scattering amplitude is studied.
Section \ref{sec:Large-E-and-GDH} contains a discussion of the large
energy behavior of the spin dependent covariant amplitudes. The
possibility of modifying GDH sum rule is revealed. It is due to a
localized spontaneous breaking down of the EM gauge symmetry inside a
nucleon, which makes it superconducting.  The propagation of a photon
inside a nucleon is discussed in section \ref{sec:gamma-propagation},
where the resonant response of a nucleon, which contains a
superconducting phase, is studied. It also provides more concrete
basis for the discussions in section
\ref{sec:Large-E-and-GDH}. Section \ref{sec:Summ} contains discussions
and a summary. In the appendix, a brief summary of some of the recent
developments on the subject.

\section{Gauge invariance and the structure of the EM current-current
       correlator}
\label{sec:G-inv}

   The photon--nucleon forward Compton scattering amplitude is
related to the following covariant one photon irreducible
correlator between EM current operator $J^\mu$
\begin{eqnarray}
     T^{\mu\nu}(p,q) &=& i\int d^4x e^{i q\cdot x}
                 \bra{pS'} T^* J^\mu(x) J^\nu(0)\ket{pS},\label{Correlator1}
\end{eqnarray}
where $\ket{pS}$ is a nucleon state with 4-momentum $p^\mu$,
polarization $S_\mu$ and $T^*$ represents time ordering with proper
Schwinger terms added. Taking into account of the conservation of
parity in the EM interaction, the amplitude can be parameterized by
four invariant amplitudes $S_{1,2}$ and $A_{1,2}$ in the following way
\begin{eqnarray}
    T^{\mu\nu}(p,q) &=&  S_1\left (-g^{\mu\nu} + {q^\mu q^\nu \over q^2} \right
)
                       + S_2 \left (p^\mu-{m\nu\over q^2} q^\mu\right)
                             \left (p^\nu-{m\nu\over q^2}
q^\nu\right)\nonumber\\
                    && - i A_1 \epsilon^{\mu\nu\alpha\beta}q_\alpha S_\beta
                       - i m \nu
                         A_2 \epsilon^{\mu\nu\alpha\beta}q_\alpha
                         \left (S_\beta - {S\cdot q \over m\nu} p_\beta \right
)
\label{Amplitude1}
\end{eqnarray}
since the amplitude $T^{\mu\nu}$ satisfies the Ward identity
\begin{eqnarray}
     q_\mu T^{\mu\nu}(p,q) &=& 0 \label{T-ward}
\end{eqnarray}
due to the gauge invariance and commutativity of the EM charge density
operator at equal-time. Here m is the mass of a nucleon and
$\nu = p\cdot q/m$.

The simple form of the above expression is not a suitable one for us to
discuss the constraint of the gauge invariance, the natural one is
\begin{eqnarray}
     T^{\mu\nu}(p,q) &=& {1\over 2 m}
                      \bar U(pS)\left [F_1 g^{\mu\nu} - F_2 q^\mu q^\nu + F_3
                       p^\mu p^\nu - F_4 \left (p^\mu q^\nu + p^\nu q^\mu
                       \right )
                       + i F_5 \sigma^{\mu\nu}\right .  \nonumber \\ &&
                       + \left .
                            i F_6 \left (p^\mu\sigma^{\nu\alpha} q_\alpha -
                            p^\nu\sigma^{\mu\alpha} q_\alpha \right )
                       + i F_7 \left (q^\mu \sigma^{\nu\alpha} q_\alpha -
                            q^\nu\sigma^{\mu\alpha} q_\alpha \right )
                       + i F_8\epsilon^{\mu\nu\alpha\beta} q_\alpha
                       p_\beta {\rlap\slash q}\gamma^5
                          \right ] U(pS),
\label{Amplitude2}
\end{eqnarray}
where $U(pS)$ is the nucleon spinor,
$F_{1,\ldots,8}$ are invariant amplitudes that are otherwise independent
if gauge invariance is not required. The Ward identity Eq.
\ref{T-ward} imposes
the following constraints on $F_i$ $(i=1,\ldots,8)$, namely,
\begin{eqnarray}
       F_1(q^2,\nu) - q^2 F_2(q^2,\nu) - m \nu F_4(q^2,\nu)
                                        & = & 0, \label{G-inv1}\\
       m \nu F_3(q^2,\nu) - q^2 F_4(q^2,\nu) & = & 0, \label{G-inv2}\\
       F_5(q^2,\nu) - m\nu F_6(q^2,\nu) - q^2 F_7(q^2,\nu)
                                 & = & 0. \label{G-inv3}
\end{eqnarray}
  Eqs. \ref{G-inv1}--\ref{G-inv3} are regarded as dynamical constraints
on invariant amplitudes $F_{1,\ldots,8}$, each one of which
contains contributions from different parts for the EM
current--current correlator. The
reduced form, Eq. \ref{Amplitude1}, tends to hide the origin of these
amplitudes  despite its simplicity.

The reduced set of invariant amplitudes $S_{1,2}$ and $A_{1,2}$ are
related to $F_{1,\ldots,8}$ through
\begin{eqnarray}
     S_1(q^2,\nu) &=& -F_1(q^2,\nu) = -q^2 F_2(q^2,\nu) - m \nu F_4(q^2,\nu)
     ,\label{S1-F}\\
     S_2(q^2,\nu) &=& F_3(q^2,\nu)
                  = {q^2\over m \nu} F_4(q^2,\nu),\label{S2-F}\\
     A_1(q^2,\nu) &=& m F_6(q^2,\nu) + \nu F_8(q^2,\nu),\label{A1-F}\\
     A_2(q^2,\nu) &=& {1 \over m} F_7(q^2,\nu) + F_8(q^2,\nu).\label{A2-F}
\end{eqnarray}

  From crossing symmetry, which requires $T^{\mu\nu}(p,q) =
T^{\nu\mu}(p,-q)$, the following constraints
\begin{eqnarray}
      F_{1,\ldots,3}(q^2,\nu) = F_{1,\ldots,3}(q^2,-\nu), &\hspace{20pt} &
      F_4(q^2,\nu)    = - F_4(q^2,-\nu),\label{Cross1}\\
      F_{5,7,8}(q^2,\nu) = - F_{5,7,8}(q^2,-\nu), &\hspace{20pt} &
      F_6(q^2,\nu)       = F_6(q^2,-\nu)\label{Cross2}
\end{eqnarray}
have to be satisfied by $F_{1,\ldots,8}$.

\section{Large energy behavior and the possibility of modifying the
          GDH sum rule}
\label{sec:Large-E-and-GDH}

   The relation between $F_5(q^2,\nu)$, $F_6(q^2,\nu)$ and
$F_7(q^2,\nu)$ imposed by gauge invariance is given by Eq.
\ref{G-inv3}. The large $\nu$ behavior of $F_5$ can be obtained quite model
independently by using the large $\nu$ behavior of the naive time
ordered product of two currents.

   Let us consider the following naive time-ordered correlator
\begin{eqnarray}
    \tilde T^{\mu\nu}(p,q) &=& i\int d^4x e^{i q\cdot x}
                 \bra{pS'} T J^\mu(x) J^\nu(0)\ket{pS}.\label{Correlator2}
\end{eqnarray}
It has the following asymptotic behavior as $q_0\to\infty$ and ${\bf
  q}$ kept fixed \cite{BJORK}
\begin{eqnarray}
  \lim_{q_0\to \infty} \tilde T^{\mu\nu}(p,q) &=& {1\over q_0}\int d^3x
              e^{-i{\bf q}\cdot {\bf x}}
              \bra{pS'}[ J^\mu(0,{\bf x}), J^\nu(0,0)] \ket{pS}
      + O({1\over q_0^2}).\label{Asymp1}
\end{eqnarray}
In the rest frame for a nucleon, $\nu = q_0$. So, in that frame,
\begin{eqnarray}
  \lim_{\nu \to \infty} \tilde T^{\mu\nu}(p,q) &=& {1\over \nu}\int d^3x
              e^{-i{\bf q}\cdot {\bf x}}
              \bra{pS'}[ J^\mu(0,{\bf x}), J^\nu(0,0)] \ket{pS}.\label{Asymp2}
\end{eqnarray}
Neither Eq. \ref{Correlator2} nor the commutator in Eqs. \ref{Asymp1}
and \ref{Asymp2} are in general covariant under Lorentz
transformation. They contain Schwinger terms that depend on the
spatial components of $q^\mu$ and break the Lorentz
covariance. Therefore Eq. \ref{Asymp2} can not be applied to any frame
in general. However, for the covariant parts of Eqs.
\ref{Correlator2}--\ref{Asymp2}, Eq. \ref{Asymp2} is correct in any
frame. These parts are the ones that are of physical
interest. Eq. \ref{Asymp1} can in fact be generalized to the case in
which $q^2$ is fixed and $\nu$ is let to go to infinity
\cite{AdlerBook} by using an infinite momentum frame in which a sum
rule can be obtained for the commutator of the current (GDH sum rule
is assumed valid in this paper).  Using the form of the EM current
operator written in terms of quark fields, we can write down the naive
commutator between the currents {\em in a specific frame}, namely
\begin{eqnarray}
    [J^\mu(0,{\bf x}), J^\nu(0,0)] =  - 2i\epsilon^{0\mu\nu\lambda}
                 \bar \psi(0)\gamma^\lambda\gamma^5Q^2\psi(0) 
                 \delta^{(3)}({\bf x})+
                       \mbox{Schwinger terms},\label{Commutator1}
\end{eqnarray}
where $Q$ is the charge operator. The naive equal-time commutator in
Eq. \ref{Commutator1} is not Lorentz covariant. The canonical value of
the commutator is defined here to coincide with the naive one in
Eq. \ref{Commutator1} when its matrix elements are taken between
nucleon states at rest. So the matrix element of the canonial
equal-time commutator between two EM current density operators can be
obtained from the rest frame matrix elements by a proper Lorentz 
transformation. It is of the following form
\begin{eqnarray}
  \bra{pS'}\bar\psi(0) \left [   J^\mu(0,{\bf x}), J^\nu(0,0) 
      \right ]_{C} \psi(0) \ket{pS} &=&- i G^N_{axial} 
     \epsilon^{\mu\nu\alpha\beta} {p_\alpha\over m}  
    \bar U(pS') \gamma_\beta\gamma^5 U(pS) =
    G^N_{axial} \bar U(pS') i\sigma^{\mu\nu} U(pS), \label{tensorCharge}
\end{eqnarray}
with $p_\mu$ the nucleon 4-momentum and $G^N_{axial}$ the finite
``charge square weighted axial charge'' of a nucleon.

Then the form of Eq. \ref{Asymp2} in an infinit momentum frame of the
nucleon takes the form 
\begin{eqnarray}
  \lim_{\begin{array}{c}\nu\to\infty\\q^2=\mbox{fixed}\end{array}}
            \tilde T^{\mu\nu}(p,q)
            &=&
            \lim_{\begin{array}{c}\nu\to\infty\\q^2=\mbox{fixed}\end{array}}
                 \left ({1\over\nu}
              G^N_{axial} \bar U(pS') i\sigma^{\mu\nu} U(pS)
          + {1\over\nu} 
            \bra{pS'}\mbox{Schwinger terms} \ket{pS} \right ).\label{Limit1}
\end{eqnarray}

The difference between $T^{\mu\nu}$ and $\tilde T^{\mu\nu}$ contains
only non--covariant Schwinger type of terms, therefore
the covariant $T^{\mu\nu}$ and naive $\tilde T^{\mu\nu}$ contains the
same operator for the canonical part of the commutator term since it
is covariant and independent of $q^\mu$.
Therefore, in the large $\nu$ limit, the asymptotic $F_5$ behavior
can be identified, namely,
\begin{eqnarray}
        F_5 \sim \nu^{-1} G^N_{axial}. \label{F5asympt}
\end{eqnarray}

Eq. \ref{G-inv3} acquires more predictive power after the large $\nu$
behavior of $F_5$ is established. Since from Eq.
\ref{F5asympt}, $F_5$ term can be dropped from Eq. \ref{G-inv3} at
large $\nu$ and finite $q^2$; this leads us to
\begin{eqnarray}
  F_6 = - {q^2\over m\nu} F_7. \label{F6asmpt1}
\end{eqnarray}
It means that $F_6$ decreases faster than $\nu^{-1}$ at $q^2=0$ if $F_7$ is
finite. Actually, for a real photon and $F_7$ non-singular, $m F_6=
F_5/\nu\sim \nu^{-2}$.
It will be discussed in the following that in the large $\nu$ limit,
$F_8$ decrease as or
faster than $\nu^{\alpha-2}$. It causes no problem.
These are part of the
theoretical basis for deriving the  GDH sum rule. Its
validity is intimately related to the gauge invariance.

Are the assumptions that went into the derivation of the GDH sum rule as
solid as it sound? The answer is no. This is because $F_7$ is a
contribution from the part of the current--current correlator in which
one of the current contributes its longitudinal component. We have
learned from our understanding of the spontaneous breaking of  gauge
invariance due to Higgs mechanism that it can has
a massless pole corresponding to world-be Goldstone bosons
due to the symmetry breaking. For a gauge symmetry that is
spontaneously broken, these massless Goldstone bosons decouple from
physical spectrum. They however have effects in the physical sector of
the states. For example, their coupling to the gauge bosons generates
finite masses for certain gauge bosons. This point will be discussed in
the following section in the context of QED.

It is therefore  natural to define
the order parameter $\rho_\infty$ for the EM gauge symmetry
inside a nucleon in the following limit,
\begin{eqnarray}
        \rho_\infty
         &=& -\lim_{q^2\to 0}\lim_{\nu \to \infty} {q^2 F_7\over \nu
           }.\label{orderpara}
\end{eqnarray}
The first limit $\nu\to\infty$ is necessary since localized
spontaneous gauge symmetry breaking is discussed here. A finite
region in space only looks more and more like an infinite system when
smaller and smaller distances or higher and higher energies
are probed. The physical reason behind this statement will be discussed in the
next
section. For any finite $\nu$, it is
expected that $q^2 F_7=0$ at $q^2=0$ since the world-be Goldstone
boson corresponding to localized EM gauge symmetry breaking is
expected to be also localized inside the nucleon; it does not has any
effects at low energies. $\rho_\infty\ne 0$ means that for any
$q^2=\epsilon$, with $\epsilon$ chosen to be arbitrarily small, there
exists a lower bound $\nu_M$ for $\nu$ beyond which $q^2 F_7$ can be
made finite and independent of $\epsilon$.
Therefore, the order of the limits in Eq. \ref{orderpara} is important.

It was learned from the
hypothetical ``charged photon'' pion Compton scattering
\cite{BGLL,Singh} that isovector component of the current--current
correlator has to contain a $J=1$ fixed pole in order for the mutual
consistency between the Regge phenomenology and current algebra to
uphold. The absorptive part of the current--current correlator behaves
according to the Regge asymptotics since they relates to
the contributions from the physical hadronic states. For the realistic case of
photon--nucleon Compton scattering, there is no such a need from the
consistency point of view that there should be a $J=1$ fixed pole
since the time component of the EM current operator at equal-time
commutes with each
other so that the EM current--current correlator satisfies the Ward
identity Eq. \ref{T-ward}, which is likely to be violated if a fixed
$J=1$ pole is straight forwardly introduced in the same way as that of
Refs. \cite{BGLL,Singh}. The possible modification of the commutator
between time component of the EM current at equal-time
is proposed in Ref. \cite{KS,CLW}.
It is however remains to show how the Lorentz invariance is to be
recovered, what are the effects of a modification of the commutation
relation between
EM charge density operator on the low energy theorem \cite{LGG} that
enters the GDH sum rule \cite{KS}.

We shall assume that Eq.
\ref{T-ward} is true in this paper and try to see how to introduce an
equivalent effect consistent with gauge invariance, vanishing of
the equal-time commutator between two EM charge density operators and
Lorentz invariance summarized by Eq. \ref{T-ward}.

Now, the case of $\rho_\infty\ne 0$ is of interest since it would lead to a
modification of the GDH sum rule suggested by experimental data. If
$\rho_\infty\ne 0$, then the invariant amplitude $F_6$ contains a constant
contribution due to Eq. \ref{F6asmpt1}, namely
\begin{eqnarray}
      F_6(q^2,\nu) &\stackrel{\nu\to\infty}=& \rho(q^2,\nu)/m.\label{F6asympt2}
\end{eqnarray}

 It means that $mF_6$ asymptotically approaches to $\rho$
in the kinematic region $\nu\to\infty$, $q^2\to 0$. $\rho(q^2,\nu)$
may has
different pole and cut structure from $F_6(q^2,\nu)$
in the entire complex $\nu$ plane. Due to
the crossing symmetry, Eq. \ref{Cross2}, $\rho(q^2,\nu)=\rho(q^2,-\nu)$.
So it is reasonable to require
$\displaystyle\lim_{|\nu|\to\infty}\rho(q^2=0^+,\nu)=\rho_\infty=\mbox{const}$.
It is expected to be a slow varying function of $\nu$ at large $\nu$
and can be non-analytic in $1/\nu$ in the $\nu\to\infty$ limit.

 In order to study the large $\nu$ behavior of $A_1(q^2,\nu)$, the
large $\nu$ behavior of $F_8(q^2,\nu)$ has to be established. Since
$F_8(q^2,\nu)$ is an invariant amplitude that correspond to a piece in
$T^{\mu\nu}$ which is gauge invariant by itself, its large $\nu$
behavior is expected to follow the Regge asymptotics \cite{Regge}.
This is due to a fine point connected to a local
gauge theory, namely, it is known that in a gauge theory, the physical
states are contained in a subspace of the full Hilbert space of the system
due to the superselection rule imposed by ``Gauss law'' \cite{GSBK}. 
The states outside of this subspace
are called unphysical states. For example, the would-be Goldstone
boson for the spontaneous EM $U(1)$ symmetry breaking does not lie in
the physical subspace \cite{WBG}.
Since $F_8$ contains only gauge invariant pieces by itself and the
EM current is a scalar under the 
gauge transformation, the corresponding components of
the current only connects
physical initial or final states to physical intermediate states (due
to the generalized Wigner--Eckart theorem);
there is no need for the
cancellations between contributions of unphysical states
to ensure gauge invariance, which is needed for other $F_i$ (i=1,\ldots,7). 
Since 
it is known that physical scattering amplitudes are bounded by the
Regge asymptotics. So our assumption about the large $\nu$ behavior 
of $F_8$ is safe.
In the normal case, the asymptotic behavior of $A_1(q^2,\nu)$ is
expected to be $\nu^{\alpha-1}$ with $\alpha<1$, so $F_8(q^2,\nu)$
decreases as $\nu^{\alpha-2}$ or faster following Eq. \ref{A1-F}.
Therefore, the large $\nu$ behavior of $A_1$ in this case
is entirely determined by $F_6$,
which approaches a constant at large $\nu$, if $\rho_\infty\ne 0$.

\begin{figure}[h]
\epsfbox{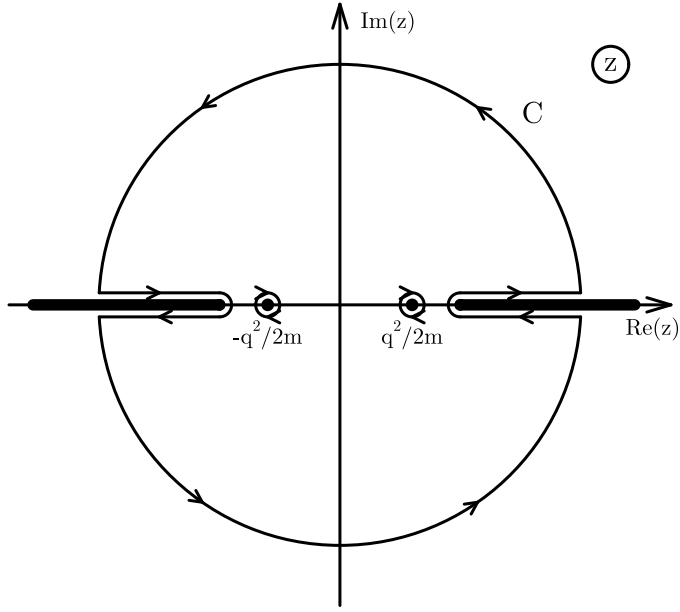}
\caption{\label{Contour} The integration contour in the complex
  $\nu\equiv z$ plane.}
\end{figure}

In this case, there is an additional term to the GDH sum rule. It can be
derived from the following expression of $A_1(q^2,\nu)$ at
sufficiently small $q^2$ and large radius $\nu_C$ for the circle $C$ in Fig.
\ref{Contour}
\begin{eqnarray}
    A_1(q^2,\nu) &=& {Res[A_1(q^2,q^2/2m)]\over \nu-q^2/2m} +
                   {Res[A_1(q^2,-q^2/2m)]\over \nu+q^2/2m} +
                 {1\over\pi}\int_{\nu_0}^{\nu_C} d\nu'
                 {ImA_1(q^2,\nu')\over \nu'-\nu}  \nonumber \\
                && \hspace{1 in}+{1\over\pi}\int_{-\nu_C}^{-\nu_0} d\nu'
                 {ImA_1(q^2,\nu')\over \nu'-\nu} + {1\over m}\int_C {dz\over
2\pi
                   i}{\rho(q^2,z)\over (z-\nu)},\label{A_1}
\end{eqnarray}
where $Res$ represents the residual of $A_1(q^2,\nu)$ on its Bonn poles
and the integration contour is given in Fig. \ref{Contour} by assuming
analyticity of $A_1(q^2,\nu)$ in the complex $\nu$ plan off the real
$\nu$ axis. $2 Im A_1(\nu) = A_1(\nu + i\epsilon)-A_1(\nu-i\epsilon)$ in Eq.
\ref{A_1}.
When the radius of the big circle $C$ in Fig. \ref{Contour} approaches
infinity, considering the crossing symmetry (Eq. \ref{Cross2}),
the above expression yield the following result of the
modification of the GDH sum rule, namely,
\begin{eqnarray}
\int_{0}^\infty {\sigma_{3/2}(\nu) - \sigma_{1/2}(\nu)\over\nu} d\nu =
 {2\pi^2\alpha\over m^2}\left (\kappa^2 + 2 m^2\rho_{\infty}\right ) ,
\label{ModGDH}
\end{eqnarray}
where the additional term comes from the large circle $C$
in the complex $\nu$ plane (see Fig.
\ref{Contour}) with $|\nu|\to\infty$.
There is no cut for the function $\rho(q^2,\nu)$ on
the real $\nu$ axis since the world-be Goldstone boson is not a
physical state and it does not couple to the physical intermediate
state (in a gauge invariant calculation) directly. Its function is to
mix with the massless photon to generate a massive excitation inside
the localized superconducting phase. This point will be
substantiated in section \ref{sec:gamma-propagation}. We therefore expect that
the absorptive part of $F_6(q^2,\nu)$ is still strongly convergent at
large $\nu$ to render an unsubtracted sum rule possible.

   This version of the GDH sum rule is effectively like the $J=1$ fixed pole
one given by  Refs. \cite{AG,FGFD} due to the same asymptotic behavior is
assumed. This paper
nevertheless discussed the
necessity of spontaneous breaking down of the EM gauge symmetry inside
a nucleon in order for this scenario to be consistent with gauge
invariance, the assumption that the time component of
the EM current operator at equal-time commutes with each other and Lorentz
invariance.  The physical origin
of the effect of the ``fixed J=1 pole'' is now clear under the above
mentioned assumptions; it is due to the
unphysical world-be Goldstone boson inside a nucleon where the EM
gauge symmetry is spontaneously broken down and where a localized
superconducting phase is formed.

   Numerically, the value for $\rho_\infty^p$ and $\rho_\infty^n$ and
their difference can be extracted from the results of the integration of
the total photon--nucleon cross sections given in Ref. \cite{SWK}. They
take the value $\rho_\infty^p = 2.9\times 10^{-2} fm^2$ and
$\rho_\infty^n= -2.5\times 10^{-2} fm^2$. As a
final remark for this section, it should be mentioned that
$\rho_\infty^{p,n}$ has a dimension of length squared.
It can be written as $\rho_\infty^{p,n}=1/\Lambda^2$. It is interesting
to note that $\Lambda\sim 1 GeV$, which correspond to the natural scale
of the spontaneous chiral symmetry breaking in strong interaction.

\section{The propagation of a photon inside a nucleon}
\label{sec:gamma-propagation}

   Let's consider the following connected correlator between the EM
gauge fields $A^\mu$
\begin{eqnarray}
   M^{\mu\nu}(p,q) &=& i\int d^4x e^{i q\cdot x}\bra{pS'}T^* A^\mu(x)
                       A^\nu(0)\ket{pS} \label{AAcorrelator}
\end{eqnarray}
in, e.g., Landau gauge. From the Lehmann--Symanzik--Zimmermann (LSZ)
reduction formalism, $T^{\mu\nu}(p,q)$ is related to
$M^{\mu\nu}(p,q)$ in the following way
\begin{eqnarray}
     T^{\mu\nu}(p,q) &=& \lim_{q^2\to 0} Z^{-2}_\gamma
                                    q^4 M^{\mu\nu}(p,q),
\label{MtoT}
\end{eqnarray}
with $Z_\gamma$ the wave function renormalization constant for a photon.
This is true because a photon outside of a nucleon at $q^2=0$ can propagates
infinite far away from the nucleon to form an asymptotic state.

   The problem that is not very well understood is how does the assumed
superconducting phase inside a nucleon affect the behavior of
a photon propagating in the same nucleon before the photon interacts with
valence quarks.  Apparently the method used in an
infinite system has to be modified when dealing with a system with
finite spatial dimension. In order to proceed, let's first write down the
formal Schwinger--Dyson equation for the dressed propagator $G$ of a photon
inside an infinite medium, namely
\begin{eqnarray}
       G &=& G_0 + G_0 \Pi G, \label{ph-S-D-eq}
\end{eqnarray}
where $G_0$ is the free propagator for a photon, $\Pi$ is the proper
self-energy of a photon inside the medium and all spin indices are
suppressed. It has the following formal solution,
\begin{eqnarray}
       G &=&  G_0 {1\over 1- \Pi G_0}. \label{G-solution}
\end{eqnarray}
The pole position is at $\Pi G_0=1$ when the pole position of $G_0$
and $G$ differs. The situation in which the pole of $G_0$ coincide
with that of $G$ is not considered in this paper.
Eq. \ref{G-solution}
can be expanded as an infinite series in certain region of $\Pi G_0$
\begin{eqnarray}
      G &=& G_0 (1 + \Pi G_0 + \Pi G_0 \Pi G_0 +
        \ldots + \underbrace{\Pi G_0 \ldots \Pi
          G_0}_{\mbox{n $G_0$s}}+\ldots ).
         \label{G-expansion}
\end{eqnarray}
Then, use can be made of the complex continuation to extend the result. This
form is the proper one for us to discuss the propagation of a
photon inside a nucleon.

Before considering a finite system in which the EM gauge symmetry is
spontaneously broken down, it is useful to review the Higgs
mechanism in the present context. For QED, the self energy of a
photon with explicit indices inserted can be written as
\begin{eqnarray}
      \pi^{\mu\nu}(q) = (q^2 g^{\mu\nu} - q^\mu q^\nu) \pi(q^2)
\label{QEDpself}
\end{eqnarray}
due to gauge invariance expressed as $q_\mu\pi^{\mu\nu} =0$, similar to
Eq. \ref{T-ward}. The spontaneous breaking down of the EM gauge
symmetry manifests itself by having a massless pole in $\pi(q^2)$.
In this case, $\pi(q^2)$ can be written
as $\pi(q^2) = m^2_\gamma/q^2 + ...$ and the ``...'' terms can be
ignored for our purposes. The free photon propagator in Landau gauge is
\begin{eqnarray}
     G_0^{\mu\nu} &=& \left (g^{\mu\nu}-{q^\mu q^\nu \over
                                   q^2+i\epsilon}\right ) {-i\over q^2 +
                                 i\epsilon}.
\label{G-0-photon}
\end{eqnarray}
Substituting Eqs. \ref{QEDpself} and \ref{G-0-photon} into Eq.
\ref{G-expansion}, one find the following expression for the
component of the full propagator induced by the transverse
$G_0^{\mu\nu}$, namely,
\begin{eqnarray}
      G_{TT}^{\mu\nu} &=& \left (g^{\mu\nu}-{ q^\mu q^\nu \over
                                    q^2+i\epsilon} \right )
                                {-i  \over q^2 - m_\gamma^2+
                                 i\epsilon},
\label{G-photon1}
\end{eqnarray}
which has a pole at $q^2=m^2_\gamma$. The pole at $q^2=0$ is still
present. In fact the pole at $q^2=0$ does not exist in the physical
excitation. This is because the
propagator in Eq. \ref{G-photon1} contains only the transverse
excitation, the longitudinal component of the same excitation can be
obtained by replacing the $G_0$ in Eq. \ref{G-expansion} by the
would-be Goldstone boson in $\pi(q^2)$ and the self-energy $\Pi$ in
the same equation by the free photon propagator coupled to the
would-be Goldstone boson. The result is
\begin{eqnarray}
  G^{\mu\nu}_{LL} = i{q^\mu q^\nu\over m^2_\gamma
(q^2+i\epsilon)}.\label{G-photon2}
\end{eqnarray}
So the full propagator of the excitation is
\begin{eqnarray}
      G^{\mu\nu} &=& G_{TT}^{\mu\nu} + G_{LL}^{\mu\nu}
                =  \left (g^{\mu\nu}-{q^\mu q^\nu \over
                                     m^2_\gamma} \right )
                                {-i  \over q^2 - m_\gamma^2+
                                 i\epsilon}.
\label{G-photon}
\end{eqnarray}
It corresponds to a massive excitation with mass $m_\gamma$. This
causes the Meissner effect in such a system, which is also called
superconducting.

The essential features of the photon propagation inside a finite
system can be learned by considering
a spherical cavity of radius $R$ in which $\Pi$ is
significantly different from the outside vacuum.
A space--time description is beneficial in this case.
The time $\Delta t$ a photon spend to
propagate through a nucleon is finite. In case of non geometric
resonance\footnote{The kind of resonance formed due to coherent multiple
                reflections (or scattering) of a photon on the
                surface of the cavity.},
it is $\Delta t = 2 R/c$ with $c=1$ the speed of light. For a
geometric resonance case, a photon might be scattered $n_G$ times
before it leaves the cavity. If the cavity is of infinite radius, the
pole of $G$ is located at $\Pi G_0=1$. When the size of the the cavity
is reduced to a finite value, the condition $\Pi G_0=1$ renders the
coherent parts of the
infinite series Eq. \ref{G-expansion} to become effectively a
finite one. In order to see this, we shall recall that the
quantity $\Pi G_0 = z(\delta t)$ has a phase factor
$e^{-i\omega\delta t}$ with $\omega$ the frequency of the photon and
$\delta t$ the time interval between two interactions of the
free photon with the medium. The condition $\Pi G_0=1$ requires
that $\omega \delta t = 2 k\pi$ ($k=\pm 1,\pm 2,\ldots$). So, at
the resonance point, which, in an infinite system,
corresponds to at the pole position of $G$,
the maximum number of interactions between a photon and the medium
inside the cavity is
\begin{eqnarray}
N_{\mbox{\footnotesize max}} &=& {\omega\Delta t\over 2 \pi} = {n_G \omega
R\over \pi},
\label{N_max}
\end{eqnarray}
where $c=1$ has been taken. So, in the region $|\Pi G_0|<1$, the
coherent part of the photon propagator can be written as
\begin{eqnarray}
  G &\approx&  G_0 (1 + \Pi G_0 + \Pi G_0 \Pi G_0 +
        \ldots + \underbrace{\Pi G_0 \ldots \Pi
          G_0}_{\mbox{$N_{\mbox{\footnotesize max}}$ $G_0$s}})\nonumber\\
      &=& G_0 {1-(\Pi G_0)^{N_{\mbox{\footnotesize max}}+1}\over 1-\Pi G_0}
.\label{finit-G-expan}
\end{eqnarray}
It has no pole at $\Pi G_0=1$ but turns to $(N_{\mbox{\footnotesize max}}+1)
G_0$ at that
point. When $N_{\mbox{\footnotesize max}}$ or $\omega R$ is let to go
to $\infty$, it indeed
becomes a pole. So, for a finite system, the dynamical resonance peak (due
to the interaction with the medium) becomes higher and higher as
$\omega\equiv \nu$ becomes larger and larger.
In mathematical language, it can be expressed as that for any finite
small deviation of $q^2$ from the pole value $m_\gamma^2$, namely
$q^2-m_\gamma^2=\epsilon$, there exist
a value for $\nu$ beyond which $(q^2-m_\gamma^2)G$ can be made finite
and independent of $\epsilon$.
At the $\nu\to\infty$
point, it turns into a pole.

This argument can also be applied
to the case of the would-be Goldstone boson discussed in the previous
section. It forms the basis for the definition given by Eq.
\ref{orderpara} and the discussions following it.

Since $M^{\mu\nu}$ defined in Eq. \ref{AAcorrelator} enters in any
forward EM scattering of a charged system with a nucleon, it can be
measured in variety of ways. The $q^2>0$ region is of interest for
the determination of the above mentioned resonant phenomenon of a
superconducting nucleon. The $q^2$ position of the resonant peak, if
exists, correspond to the mass of the photon if the size of the
nucleon is let to go to infinity. In addition, the hight of the peak
would be increased if the energy (in the lab. frame of the nucleon) of
the photon is increased. We shall call this phenomenon virtual photon
transparency at high energies.

Since a forward scattering experiment is hard
to perform, one can use the optical theorem to relate the imaginary
part of the forward
scattering amplitude to a corresponding total cross section. For
example, the total cross section of the reaction
\begin{equation}
\gamma^*_{e^+ e^-} + N \to X,\label{inclusiveA}
\end{equation}
where $\gamma^*_{e^+e^-}$ is a virtual photon with $q^2>0$
originated from an annihilation of electron and positron pair and $X$
denotes any allowed final states,
corresponds to the imaginary part of forward scattering
\begin{equation}
\gamma^*_{e^+e^-}+N\to
\gamma^*_{e^+e^-}+N.\label{forwardA}
\end{equation}
If a nucleon is superconducting, then there should be a resonant
extreme at $q^2=m_\gamma^2$ in
the real part of the forward scattering amplitude corresponding to Eq.
\ref{forwardA}. It can be shown that the imaginary part of the same
amplitude is minimized at the same point. This causes a reduction  of the
total cross section or particle production for reaction Eq.
\ref{inclusiveA}. Contrary to
intuition, the higher the virtual photon energy,
the better the relative transparency.

Finally, it should be made clear that all of the statements for a nucleon
in this section apply to any finite system, including, e.g., a nucleus.

\section{Discussions}
\label{sec:Summ}

   In summary, it is possible to reconcile the GDH sum rule and
experimental data if a localized spontaneous breaking down of the
EM gauge symmetry inside a nucleon is assumed. Complete experimental 
measurements should enable us to determine the value of
$\rho_\infty^p$ and $\rho_\infty^n$. The extension for the GDH sum rule given
in this study satisfies the following assumptions, namely, 1)
gauge invariance, 2) the assumption of the commutativity of the EM
charge density operator at equal-time and 3) Lorentz invariance.
These assumptions impose constraints so stringent that it allows very 
little freedom
for us to formulate an extension of the GDH sum rule that can be used
to explain the data if a violation of the GDH sum rule is finally
established. 

  The result of this paper is based on the truth of Eq. \ref{T-ward}, which
is a consequence of our basic assumptions mentioned above, plus a not
so explicitly stated assumption that the use of an infinite
momentum frame in deriving the fixed $q^2$ sum rule is legitimate. Without
assuming the spontaneous breaking down of the $U(1)$ EM gauge
symmetry inside a nucleon, a violation of the GDH sum rule would imply (see also Ref.
\cite{FGFD}) either the need for a modification of the commutation
relation between EM charge density operator, which has already 
been attempted in Refs. \cite{KS,CLW}, or the use of the
infinite momentum frame in the derivation of 
fixed $q^2$ sum rule being invalid
for the study of the GDH sum rule.
The first alternative requires quite drastic
changes since the Ward identity Eq. \ref{T-ward} has been assumed in 
almost all calculations involving real- and
virtual- photon interaction with a nucleon. 
The later possibility has not been explored in a constructive way. It
is not expected to be generally true given the success of other fixed
$q^2$ sum rules. How can it fail in deriving GDH sum rule 
is worthy of study in the future.  

Of course the Regge asymptotic behavior argument for $F_8$ can also
be false. If it is indeed the case and $F_8\sim \nu^{-1}$, then a
violation of the GDH sum rule can also be realized without the need of
a violation of the Ward identity Eq. \ref{T-ward}. This possibility 
is a natural alternative to the one proposed here. What can
cause such an asymptotic behavior for $F_8$, which contains the pieces
in the current--current correlation that is gauge invariant by
themselves, is a worthy topic to be further investigated. It is however
a quite unlikely scenario since, as it is discussed in the section
\ref{sec:Large-E-and-GDH}, it violates the requirements of a
superselection rule for physical states
in the EM gauge theory if the Regge asymptotics
is assumed to be correct for physical amplitudes.

  What has not been answered in this paper is how can a would-be
Goldstone boson corresponding to EM gauge symmetry be dynamically
generated in a hadronic system. One possibility was given in Ref.
\cite{Ying1}, which studied the formation of the $\beta$--phase
(superconducting) 
inside a infinitely large massless fermion system possessing chiral 
$SU(2)_L\times SU(2)_R$ symmetry. Albeit a
non-scalar condensation of diquark is studied in Ref. \cite{Ying1},
the qualitative features given in section \ref{sec:gamma-propagation}
remain true. The quantitative result of section \ref{sec:Large-E-and-GDH} also
applies to the situation of non-scalar diquark condensation discussed
in Ref. \cite{Ying1} for a class of specific configuration of the diquark
condensate in the nucleon. The photon mass $m_\gamma$ is
estimated to be of order 10 $MeV$ in that reference if the spontaneous
chiral symmetry breaking scale is chosen to be around 1 $GeV$.
The possibility of its
existence inside a nucleon was investigated in Ref. \cite{Ying2} from
chiral symmetry point of view. It was found that empirical facts seem to
favor a superconducting nucleon\footnote{If the deviation of GDH
sum rule from the experimental data is proven true, it could provide
additional information about the distribution of diquark condensate in
a nucleon under such a scenario.}. It is important to push this
idea further in future studies.

  The next important question is about the $q^2$ evolution of the sum
rule in the $q^2<0$ region, especially in the isovector sector (i.e.
when the difference between proton and neutron is taken) where, at 
large $q^2$, a sum rule due to Bjorken \cite{BJORK} exists. It agrees 
well with the experimental data \cite{Voss}. The Bjorken sum rule applies to
the case of deep inelastic scattering (DIS). The rapid transition
region between the DIS region where parton model picture applies 
and the essentially non--perturbative region probed by the GDH
sum rule is of interest and is the subject of some experiments planed.
The existence of the resonance peak in the scattering amplitude
proposed here, which is only $\sim 10$ MeV 
away from the $q^2=0$ point in the positive $q^2$ region, 
can provide a dynamical reason for such a quick transition in the
negative $q^2$ region. 
It is another one of the interesting questions that deserves to be
further investigated.

\begin{appendix}

\section{Recent progresses on the subject}

There were two main questions raised in this paper when it was finished
in 1995.  In the first one, the author posed his doubt about the
modification of the GDH sum rule by invoking non-commuting Abelian
charge density operator at equal time, the second is about the
author's uncertain about the legitimacy of the use of an infinite
momentum frame. A constructive study of the infinite momentum limit is
called for.

 There is a recent work \cite{Pantf1,Pantf2} by Pantf\"order and
collaborators, who, apart from others, studied these two problems in
the context of electroweak theory in an one loop perturbative
scheme. It was found that the effects of the anomalous term in the
equal-time charge density commutator can be canceled by an additional
term in the process of taking the infinite momentum limit. This work
therefore provides a detailed one loop theoretical check on the truth
of Eq. \ref{T-ward}, which as it was discussed in the summary part of
this work, is our the starting point.  Although this is a perturbative
proof, this part of work partially narrow down the possible
alternatives of modifying the GDH sum rule in case that there is
indeed a need from observations.

 The meaning of the ``localized'' spontaneous EM symmetry breaking
(and chiral symmetry breaking proposed in Ref. \cite{Ying2}) are
clarified in another set of papers \cite{Npap1,Npap2,Npap3}.
According to these works, especially Ref. \cite{Npap1}, if the
violation of the GDH sum rule and/or the results of Ref. \cite{Ying2}
are finally experimentally proven, then they imply either the phase of
the strong interaction vacuum is actually locally flipped to one of
the new phases proposed (also see, e.g., Ref. \cite{Wilczek}) inside a
nucleon or one of these proposed phases for the strong interaction
vacuum is close (in energy density) to the actual phase of the strong
interaction vacuum at least inside a nucleon at the present day 
conditions.

On the observational side, the recently determined sea flavor
unsymmetry inside a nucleon by the E866/NuSea collaboration
\cite{E866} are noticeably smaller than the required unsymmetry
derived from the NMC measurement \cite{Thomas} by assuming the
validity of the Gottfried sum rule. An explanation of this genuine
violation of the Gottfried sum rule was provided long ago \cite{PCAC1}
based upon the same mechanism as the one proposed in this work. The
possibility that the mechanism proposed in this work can be utilized
to provide a coherent physical picture for us to understand the
violation of the Gottfried sum rule and the seemingly ``violation of
unitarity'' (by the so called ``hard pomeron'' \cite{Landshof}) in
high energy deep inelastic scattering between leptons and nucleons in
the small Bjorken x region observed at HERA \cite{HERA1,HERA2} and in
high energy proton-proton scattering is under investigation
\cite{Invest} based upon a mechanism of spontaneous partial breaking of
local EM gauge symmetry. 

\section{A answer to the criticism of Ref [21]}

The comments of Ref. \cite{Pantf2} on this work show that its author
holds somewhat simplified understanding of this work and the
physics of spontaneous gauge symmetry breaking.

First, let us discuss the first erratum that are thought to exist in
this work.  The alleged mistake is that $F_8$ in Eq. \ref{Amplitude2}
should be absent because it is not linearly independent. This is not
true since the matrix elements between two on shell
nucleon/antinucleon states with the same {\em momentum and spin} is a
set of truncated $2\times 2$ matrix actually. But I am talk about the
Dirac $4\times 4$ matrix structure for {\em Compton amplitude} here,
it includes not only the {\em on shell} matrix elements between
nucleon/antinucleon states with the same momentum and spin, but also
those with the same momentum but {\em different spin}. Each of the
$F_i$ ($i=1,2,\ldots,8$) carry different {\em dynamical information}
of the system. In addition, $F_8$ is different from $F_6$ and $F_7$ in
that it contains no non-physical state contributions since it is a
gauge invariant piece in itself. This is discussed in length in this
paper. Did the author of Ref. \cite{Pantf2} notice that the spin
quantum numbers for the initial and final states are different in my
equations?

The confusion of Ref. \cite{Pantf2} starts from my use of the term
``canonical commutator'' which becomes synonymous to the term ``naive
commutator'', which is not manifestly covariant, there. I admit that
the term ``canonical'' may not be the best choice in this case since,
as it is apparent in the context of the discussion around it, it
really means the covariant part of it generated from the equal-time
naive commutation relation for a nucleon {\em at rest}.  Since its
matrix elements between two nucleon states are independent of $q^\mu$
and $p^\mu$, the covariant part of the time-ordered products of two
current operators and their commutators can be uniquely identified in
any frame. Therefore I can boost the rest frame equation \ref{Asymp2}
to a large momentum frame and finally let the momentum to approach to
infinity. The $F_5$ term coincides with the term derived from the
matrix elements of the naive commutator between nucleon states at
rest. Since the naive equal-time commutator between fermion fields
will lead us to the same form of current-current commutation relation
for any frame, they are not related to one another by a Lorentz
transformation. In this sense, the ``canonical commutator'' of this
work is different from the naive commutator of Ref. \cite{Pantf2}.
This is the reason that the piece of the commutator between equal-time
EM current operators given in Eq. \ref{Commutator1} is called
``canonical''. But I admit that the quantity called ``tensor charge''
of a nucleon in Ref. \cite{JGpap} should actually be axial charge of a
nucleon.  This is corrected in this version of the paper. But, non of
the results of this work is affected by this change.

There is nothing wrong with using the Bjorken-Johnson-Low (BJL)
equation \cite{BJORK,JLpap}.  The BJL equation relates the high energy
behavior of the matrix elements of a naive time-ordered amplitudes
between two operators and the equal-time commutator of the same
operators. It is a mathematical equation for all linearly independent
components on both side of the equation. The use of it is to get the
left hand side by knowing the right hand side and vice
versa\footnote{If both sides is unknown, its useless. If, however,
both sides are known, its either trivial or wrong and therefore is
almost useless.}. In the case of this work, I happen to know one
(linearly independent) piece of covariant part of the commutator,
which is unique, but do not know the high energy behavior of the
corresponding piece in the time-ordered product. But nothing is wrong
with it, isn't it?  It is not the authentic operational BJL techniques
(BJL procedure), but the formal mathematical BJL equation. The BJL
techniques defines the $q^\mu$ dependent (polynomial) parts of
commutator for physical amplitudes. It is actually a well founded
assumption.  It was applied successfully in the study of anomalies
which was experimentally tested.  

In a free theory, it can be checked straight forwardly that $F_5\sim
\nu^{-1}$ so the contribution of the {\em pair states are expected to
be suppressed or, in a less favorable case, cancellable, in the
infinite momentum limit} in the same way as the one demonstrated in
Ref \cite{Pantf2}. Therefore, according to the normal procedure, it is
safe to take the infinite momentum limit \cite{AdlerBook}. The author
of Ref. \cite{Pantf2} uses the discussion concerning $f_2$ in the
section 4.4.4 of that reference to substantiate his claims. But the
sum rules that I was talking about is an equation, adding and
subtracting the {\em same quantity} (seagull or Schwinger terms) to an
equation does not alter the content of that equation.  Is
Ref. \cite{Pantf2} suggesting that the leading $\nu$ dependence
involved in the GDH sum rule are generated by $F_5$ term so that the
troubling ``pair term'' contributions to the naive time-order product
and the equal-time commutator are in fact not get canceled (see
Ref. \cite{AdlerBook}) in the infinite momentum limit? This seems to
invalid the its main conclusion. I am puzzled and invite a new (formal
or informal) paper from its author to clarify this!

The question related to the violation of the Burkhardt--Cottingham (BC) 
sum rule. The BC sum rule was derived based upon a superconvergence
relation 
\begin{equation}
\lim_{\nu\to\infty}\nu A_2(\nu,q^2)=0
\label{BCsum}
\end{equation} 
with an implicit assumption that the EM gauge symmetry is
unbroken. But I am assuming that the EM gauge symmetry is
spontaneously broken here! Eq. \ref{BCsum} is not experimentally
direct testable, so no one know whether it is correct or not at the
present. But the current potentially testable version of the BC sum
rule, namely,
\begin{equation}
  \int^1_0 dx g_2(x) = 0
\label{BCsum1}
\end{equation}
is not expected to be essentially altered by an infinite amount 
in the deeply inelastic
region since Eq. \ref{BCsum1} is an equation for the imaginary part of
$A_2$ which suppresses the contributions from the unphysical states
in the spontaneously broken phase of EM.

Finally about the question of whether $q^2\to 0$ should be taken or
$q^2=0$ should be taken. From a mathematical sense, the set of points
that makes $I_1(0)= - \kappa^2/4$ denoted by $q^2=0$ in
Ref. \cite{Pantf2} has a zero measure. This mathematical isolated
point is therefore unphysical and can be removed from the physical
$q^2$ axis.  As the author of \cite{Pantf2} knows, he never use any
quantity with strict $q^2=0$ in the above sense in his finite momentum
sum rule! In fact his $q^2=\nu^2/P^2_0$ which approaches to zero in
the {\em infinite momentum limit}.

     I would not say anything again about the degenerate massless state
with photon. I have discussed this in length in the paper.

 In conclusion, the relevant theoretical reasons provided in
Ref. \cite{Pantf2} in support of its author's disbelief of the
modifications of GDH sum rule proposed in this work are either
inconsistent with the ``good part'' of this reference, which actually
demonstrated the cancellation of pair state contributions to the
GDH sum rule in a perturbative way using the standard electroweak model,  
or is originated from a misunderstanding of the issue.

\end{appendix}

\end{document}